\documentclass[
reprint,
superscriptaddress,
showpacs,
amsmath,
amssymb,
aps,
pra,
nofootinbib,
floatfix
]{revtex4-1}

\usepackage{graphicx} 
\usepackage{dcolumn} 
\usepackage{bm} 

\usepackage[colorlinks=true,urlcolor=blue,citecolor=blue,linkcolor=red]{hyperref}

\usepackage{color}
\usepackage{url}
\usepackage{braket}
\usepackage{subcaption}
\usepackage{qcircuit}

\newcommand{\figref}[1]{Fig.~\ref{#1}}

\usepackage[english]{babel}
\usepackage{letltxmacro}

\LetLtxMacro{\ORIGselectlanguage}{\selectlanguage}
\makeatletter
\DeclareRobustCommand{\selectlanguage}[1]{%
  \@ifundefined{alias@\string#1}
    {\ORIGselectlanguage{#1}}
    {\begingroup\edef\x{\endgroup
       \noexpand\ORIGselectlanguage{\@nameuse{alias@#1}}}\x}%
}
\newcommand{\definelanguagealias}[2]{%
  \@namedef{alias@#1}{#2}%
}
\makeatother

\definelanguagealias{en}{english}

\begin{document}

\title{A blueprint for fault-tolerant quantum computation with Rydberg atoms}

\author{James M.~Auger}
\email{james.auger.09@ucl.ac.uk}
\affiliation{
 Department of Physics and Astronomy,
 University College London,
 Gower Street,
 London,
 WC1E 6BT,
 UK
}

\author{Silvia Bergamini}
\affiliation{
 School of Physical Sciences,
 The Open University,
 Milton Keynes,
 MK7 6AA,
 UK
}

\author{Dan E.~Browne}
\affiliation{
 Department of Physics and Astronomy,
 University College London,
 Gower Street,
 London,
 WC1E 6BT,
 UK
}

\begin{abstract}
We present a blueprint for building a fault-tolerant universal quantum computer with Rydberg atoms. Our scheme, which is based on the surface code, uses individually-addressable optically-trapped atoms as qubits and exploits electromagnetically induced transparency to perform the multi-qubit gates required for error correction and computation. We discuss the advantages and challenges of using Rydberg atoms to build such a quantum computer, and we perform error correction simulations to obtain an error threshold for our scheme. Our findings suggest that Rydberg atoms are a promising candidate for quantum computation, but gate fidelities need to improve before fault-tolerant universal quantum computation can be achieved.
\end{abstract}
\maketitle

\section{\label{sec:overview}Introduction}

Rydberg atoms are a promising candidate for quantum computation~\cite{Saffman2010-ea}, having desirable properties such as relatively simple entangling gates between many qubits and the ability to fit thousands of qubits into a very small footprint.

Although there has been much interest in using Rydberg atoms for quantum computation, including a recent 51-qubit quantum simulator~\cite{Bernien2017-hn}, very little work has considered the steps required to build a fault-tolerant quantum computer with Rydberg atoms. Previous work on error correction has been limited to \cite{Brion2008-iv}, which considered error correction within an ensemble of atoms representing a single qubit, \cite{Crow2016-ay}, which focused on using Rydberg atoms in measurement-free error correction schemes, and~\cite{Isenhower2011-wh}, which investigated error rates for multiple-controlled CNOT gates using Rydberg atoms. Additionally, methods for building a universal quantum computer with Rydberg atoms using a decoherence-free subspace to mitigate the effects of errors were suggested in~\cite{Brion2007-ly}.

In this work, we propose a Rydberg atom scheme for performing fault-tolerant quantum computation with the surface code~\cite{Kitaev2003-bz, Bravyi1998-kf}. The platform is a regular two-dimensional array of atoms with spacings of a few micrometers, which can be obtained using microscopic dipole trapping techniques~\cite{Bergamini2004-bf}. Ideas for a Rydberg atom based quantum simulator using the toric code were considered in \cite{Weimer2010-hf, Weimer2011-hr}, but our work goes beyond this to consider some of the steps required to build a fully-fledged universal Rydberg atom quantum computer with active error correction.

This paper is structured as follows. Sections \ref{sec:rydberg} and \ref{sec:surface_codes} provide a brief introduction to Rydberg atoms and the surface code respectively. We then describe our proposed scheme in \ref{sec:scheme}, before obtaining an error correction threshold for it in \ref{sec:error_thresholds}.

\section{\label{sec:rydberg}Quantum computation with Rydberg atoms}

Rydberg atoms are neutral atoms with one or more electrons in a highly-excited state, i.e. with principal quantum number $n \gg 1$ ---  the alkali metals, particularly rubidium and caesium, are the species of atoms most commonly used for Rydberg atom experiments due to their single valence electrons. One of their most useful features for quantum computation is the dipole blockade, which facilitates the implementation of entangling gates between multiple atoms. When two neighboring neutral atoms are in their ground states with separation $R$, the energy required to excite one of them to a particular Rydberg state $\ket{r}$ is $E_r$. However, once one atom is in its Rydberg state, the energy required to excite the neighboring atom to the state $\ket{r}$ is increased to $E_r + V(R)$; exciting one atom to its Rydberg state effectively \emph{blockades} the other.

The size of this energy shift $V(R)$ generally falls into one of two regimes: when the atoms are sufficiently close, the dominant interaction is due to the dipole-dipole interaction, which scales as $V(R) \sim 1/R^3$. When the atoms are sufficiently distant from each other, the dominant interaction becomes the Van der Waals interaction, which scales as $V(R) \sim 1/R^6$. This work will favor the Van der Waals regime due to the faster decay in interaction strength, which will reduce unwanted interactions between distant atoms.

Experimentally, single-qubit gate fidelities in excess of 99\% have been demonstrated~\cite{Xia2015-dm, Wang2016-tq}, but two-qubit gates are languishing behind, with the best experiments achieving fidelities of around $80\%$ when post-selecting for qubit loss~\cite{Jau2015-tf, Maller2015-ye}; it is to be noted that this is due to technical limitations rather than a fundamental limit. For a recent summary of the state of Rydberg atom experiments, we direct the reader to \cite{Saffman2016-je}.

\section{\label{sec:surface_codes}The surface code}

The surface code is a topological quantum error correction code with qubits arranged on a two-dimensional manifold. The toric code~\cite{Kitaev2003-bz} is the prototypical example and has the qubits arranged in a square lattice on the surface of a torus; this idea was later extended to the planar code, which is defined on a flat surface with boundaries~\cite{Bravyi1998-kf}. The surface code can be readily adapted to perform universal quantum computation through techniques such as lattice surgery~\cite{Horsman2012-wh} or braiding defects~\cite{Fowler2012-yi}. Throughout this work, the logical state of the surface code will be denoted by $\ket{\overline{\psi}}$.

The surface code is a type of stabilizer code, which are codes that detect errors by measuring a set of carefully chosen commuting Pauli operators known as \emph{stabilizer generators}; these stabilizer generators and their products form an abelian group known as the stabilizer, $\mathcal{S}$. When there are no errors, all operators in the stabilizer have measurement outcome $+1$, i.e.
\begin{equation*}
    S \ket{\overline{\psi}} = \ket{\overline{\psi}} \quad \forall \quad S \in \mathcal{S}.
\end{equation*}
If an error occurs that anticommutes with one or more stabilizer operations, the outcomes of these measurements will then be $-1$; the measurement outcomes provide a syndrome that can be used to diagnose the error. Arbitrary trace-preserving qubit errors collapse into Pauli errors when the stabilizer generators are measured~\cite{Nielsen2000-nv}, so it suffices to detect and correct only Pauli $X$ and Pauli $Z$ errors. Non-trace-preserving leakage errors are discussed later.

\begin{figure}[h]
    \centering
    \begin{subfigure}{0.45\linewidth}
        \includegraphics[width=\textwidth]{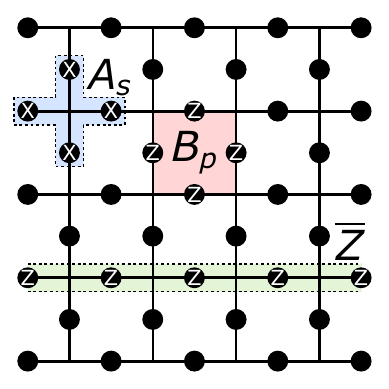}
        \caption{\label{fig:primal_lattice}}
    \end{subfigure}
    \qquad
    \begin{subfigure}{0.45\linewidth}
        \includegraphics[width=\textwidth]{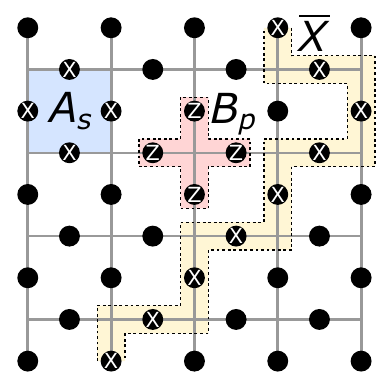}
        \caption{\label{fig:dual_lattice}}
    \end{subfigure}
    \caption{\label{fig:planar_lattice}A distance $d=5$ planar code (black dots represent physical data qubits). (\subref{fig:primal_lattice}) Primal lattice showing a star stabilizer generator $A_s$, a plaquette stabilizer generator $B_p$ and a logical Pauli operator $\overline{Z}$. (\subref{fig:dual_lattice}) Dual lattice showing the same stabilizer generators and a logical Pauli operator $\overline{X}$ that anticommutes with the logical $\overline{Z}$ operator in (\subref{fig:primal_lattice}).}
\end{figure}

\figref{fig:primal_lattice} shows the arrangement of physical \emph{data} qubits in the planar code. Each vertex, $s$,  is associated with a \emph{star} stabilizer generator, $A_s$, on the qubits surrounding the vertex
\begin{equation*}
    A_s = \bigotimes_{i \in s} X_i,
\end{equation*}
and each square, $p$, is associated with a \emph{plaquette} stabilizer generator, $B_p$, on the qubits surrounding the square
\begin{equation*}
    B_p = \bigotimes_{i \in p} Z_i.
\end{equation*}
All stabilizer generators involve only neighboring qubits.

\figref{fig:dual_lattice} shows the same lattice as \figref{fig:primal_lattice}, but with the role of stars and plaquettes swapped between vertices and squares --- this configuration is known as the dual lattice, and that in \figref{fig:primal_lattice} is known as the primal lattice. This section will discuss the detection and correction of Pauli $Z$ errors on the primal lattice, but an analogous procedure can be used to detect and correct Pauli $X$ errors on the dual lattice.

In addition to the stabilizer generators, one also defines logical Pauli $X$ and Pauli $Z$ operators, $\overline{X}$ and $\overline{Z}$. These operators commute with all star and plaquette operators, but anticommute with each other, and they are formed by strings of qubit operators between opposing boundaries of the lattice, as shown in \figref{fig:planar_lattice}. Note that the logical operators are not unique: each logical operator can be modified by multiplication with an element of the stabilizer to form a new logical operator that has the same effect on the logical state, i.e.
\begin{equation*}
 \left.\begin{aligned}
        \overline{X} \ket{\overline{\psi}} = S \overline{X} \ket{\overline{\psi}}\\
        \overline{Z} \ket{\overline{\psi}} = S \overline{Z} \ket{\overline{\psi}}
       \end{aligned}
 \right\}
 \quad \forall \quad S \in \mathcal{S}.
\end{equation*}
The weight of the lowest-weight logical operator is known as the code distance, $d$. A code with distance $d$ can reliably correct errors on $\left\lfloor{(d-1)/2}\right\rfloor$ data qubits.

Whenever a Pauli $Z$ error occurs on a single qubit, it flips the parity the adjacent star operators. When $Z$ errors occur on two qubits adjacent to a single star operator, the combined error commutes with the star operator, so the parity of the star operator will remain unchanged. However, these errors will be detected by the neighboring star operators such that only the ends of strings of errors are detected and not the actual locations of the errors.

The collective outcomes of all star measurements --- the error \emph{syndrome} --- provide the locations of the ends of all strings of errors, and it is the job of the classical \emph{decoder} to find a suitable correction for the errors. If the decoder finds a correction that results in the error being exactly corrected, the correction succeeds. Additionally, any correction that results in the errors and corrections forming a contractible loop is also a successful correction, as it is equivalent to a product of plaquettes and therefore leaves the logical state unaltered. However, corrections that combine with errors to form uncontractible strings across the lattice are equivalent to logical operations and mean a logical error has occurred and the error correction has failed.

In general, the measurement of stars and plaquettes is itself subject to error, but this can be handled by repeating the measurements many times to build up a syndrome over multiple time steps; this repeated measurement means the syndrome extraction itself does not have to be fault-tolerant. Computation and error correction are performed simultaneously, such that the stabilizer generators are measured repeatedly until the computation is complete. Corrections can either be applied as they are found, or, more likely, their effect can be propagated through the computation and accounted for when the final measurement is performed.

Numerous approaches have been suggested for surface code decoders, including minimum weight perfect matching decoders~\cite{Dennis2002-de}, renormalization decoders~\cite{Duclos-Cianci2010-qd} and cellular automaton decoders~\cite{Herold2015-oc}. An optimal decoder would provide the maximum probability of successful correction, but known algorithms for such decoders are generally computationally inefficient (with certain exceptions~\cite{Bravyi2014-kg, Delfosse2017-pa}). Decoders based on minimum weight perfect matching~\cite{Dennis2002-de} are the most widely used for the surface code as they are computationally efficient and achieve relatively high error thresholds. Minimum weight perfect matching decoders work by pairing the ends of error strings in a way that minimizes the total weight of the correction strings, with each edge of the planar code having a weight assigned to it related to the probability of an error occurring at that location.

\section{\label{sec:scheme}Proposed scheme}

In this work, we propose using individually-addressable optically-trapped neutral atoms to represent qubits in a planar code, with multi-qubit gates performed by exploiting electromagnetically induced transparency (EIT) using the methods in~\cite{Muller2009-uj}. The atoms are arranged in a two-dimensional array of micron-sized traps, with atomic spacings of a few microns~\cite{Bergamini2004-bf}. This approach has several desirable features, including parallel operation, the ability to activate local interactions with large contrast as needed via laser addressing, and robustness towards interactions between target atoms provided Rabi frequencies and interaction strengths involved (i.e. Rydberg states) are chosen appropriately, as discussed in~\cite{Muller2009-uj}.

The $\ket{0}$ and $\ket{1}$ states of each physical qubit are represented by hyperfine ground states of the atoms, and Rydberg states, labeled $\ket{r}$, are used to mediate interactions. Note that atoms involved in an interaction may utilize different Rydberg states such that $\left| r \right\rangle_i$ and $\left| r \right\rangle_j$ are not necessarily the same states for atoms $i$ and $j$.

\begin{figure}[h]
    \centering
    \begin{subfigure}{0.4\linewidth}
        \includegraphics[width=\textwidth]{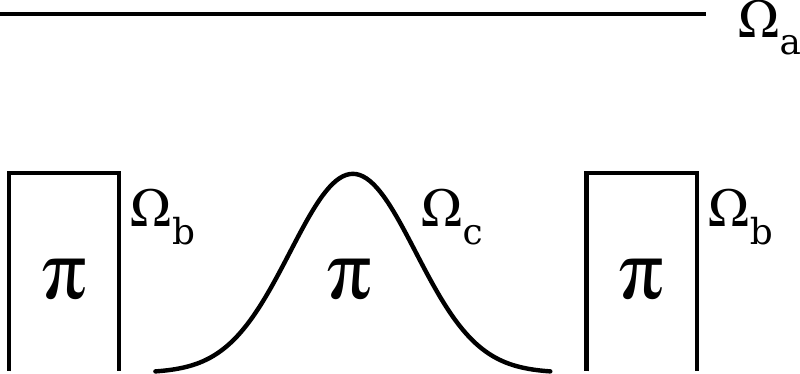}
        \caption{\label{fig:eit_pulses}}
    \end{subfigure}

    \begin{subfigure}{0.45\linewidth}
        \includegraphics[width=\textwidth]{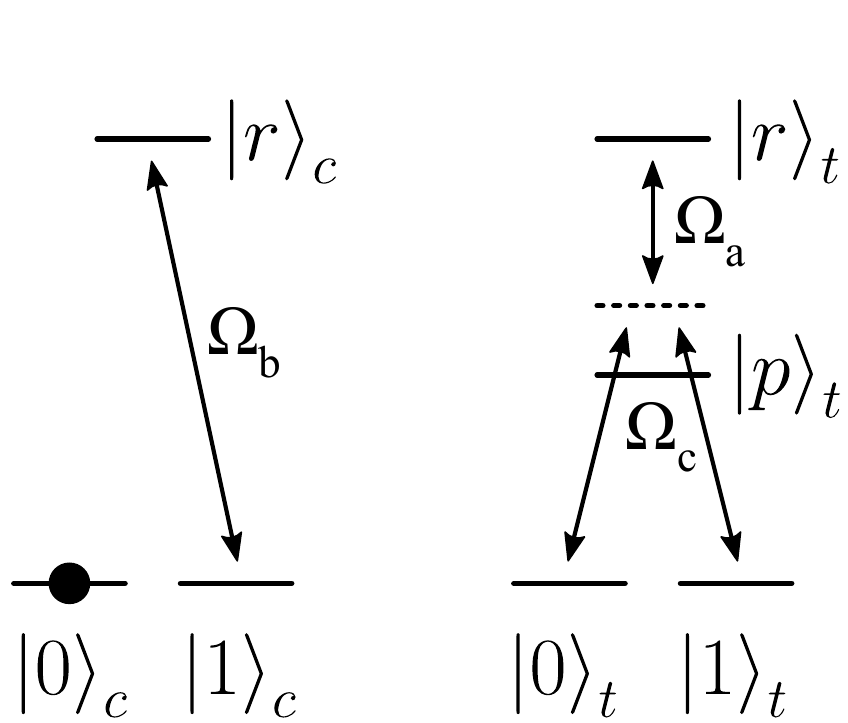}
        \caption{\label{fig:eit_blocking}}
    \end{subfigure}
    \hfill
    \begin{subfigure}{0.45\linewidth}
        \includegraphics[width=\textwidth]{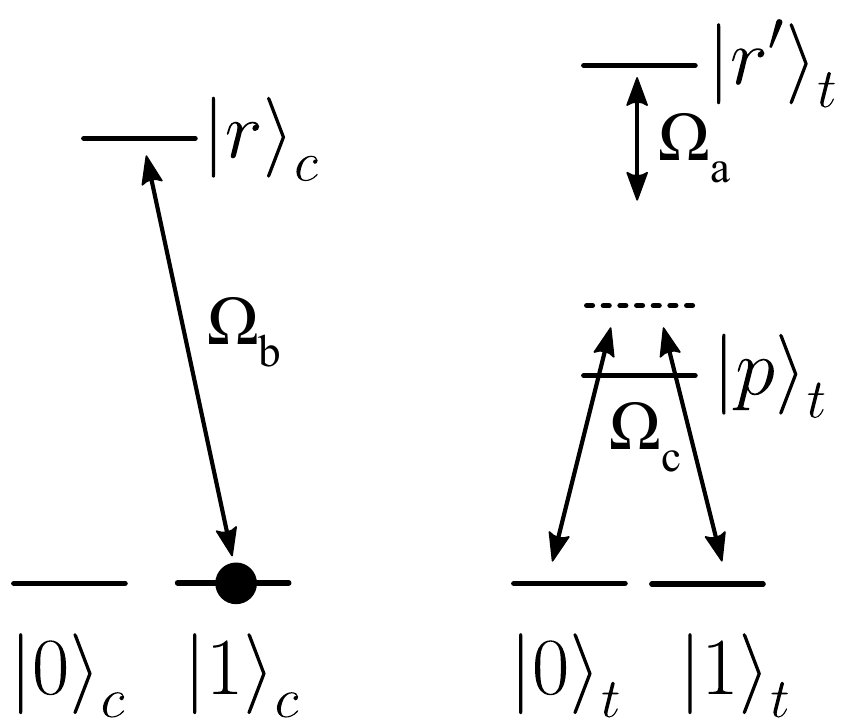}
        \caption{\label{fig:eit_not_blocking}}
    \end{subfigure}
    \caption{Using EIT to perform a CNOT gate with the method in~\cite{Muller2009-uj} (control qubit always on the left, target qubit always on the right). (\subref{fig:eit_pulses}) shows the order of the pulses, (\subref{fig:eit_blocking}) shows EIT blocking the $\ket{0}_t \leftrightarrow \ket{1}_t$ transition on the target qubit and (\subref{fig:eit_not_blocking}) shows the dipole blockade shifting the EIT out of resonance and allowing the $\ket{0}_t \leftrightarrow \ket{1}_t$ transition on the target qubit.}
    \label{fig:eit_gate}
\end{figure}

\figref{fig:eit_gate} shows the process for using EIT to perform a CNOT gate between a control and a target qubit, as proposed in~\cite{Muller2009-uj}. Initially, the $\ket{1}_c$ state of the control atom is resonantly coupled to the $\ket{r}_c$ state using a $\pi$ laser pulse with Rabi frequency $\Omega_b$. A second $\pi$ pulse with Rabi frequency $\Omega_c$ is then used to off-resonantly couple the $\ket{0}_t$ and $\ket{1}_t$ states of the target atom via an off-resonantly coupled intermediate state $\ket{p}_t$, before another $\pi$ pulse with frequency $\Omega_b$ is applied to again couple the $\ket{1}_c$ and $\ket{r}_c$ states of the control atom. Throughout this process, a strong laser with Rabi frequency $\Omega_a$, where $\Omega_a \gg \Omega_c$, is used to off-resonantly couple the Rydberg state $\ket{r}_t$ of the target to the intermediate state $\ket{p}_t$ and achieve EIT. When the control atom starts in the $\ket{0}_c$ state, the initial $\Omega_b$ pulse has no effect and the beam $\Omega_a$ prevents Raman transfer between the $\ket{0}_t$ and $\ket{1}_t$ states on the target due to EIT. When the control atom starts in the $\ket{1}_c$ state, the $\ket{r}_c$ state of the control atom becomes populated after the first $\Omega_b$ pulse, which in turn shifts the Rydberg state $\ket{r}_t \mapsto \ket{r'}_t$ of the target atom to take the $\Omega_c$ beam out of resonance and remove the EIT condition on the target, leading to an effective coupling between the $\ket{0}_t$ and $\ket{1}_t$ states.

\begin{figure}[h]
    \centering
        \includegraphics[width=0.6\columnwidth]{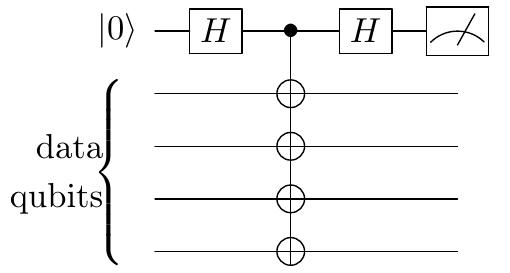}
    \caption{\label{fig:star_measurement}Measuring a star operator using a multi-target EIT CNOT gate. The top qubit is the ancilla syndrome qubit and is used only for syndrome measurement. The method for measuring a plaquette stabilizer generator involves applying Hadamard gates to each data qubit before and after the entangling operation but is otherwise identical.}
\end{figure}

This method can be used to perform simultaneous CNOT gates between a single control qubit and multiple target qubits, making it ideal for syndrome measurement in the surface code using the measurement circuit shown in \figref{fig:star_measurement} --- every star and plaquette has an associated ancilla \emph{syndrome} qubit used to measure the stabilizer generators. The $\Omega_c$ pulse is of the order of a few 10s of MHz such that the multi-qubit interaction can be performed in under a millisecond~\cite{Muller2009-uj}.

The fidelity of the process is dependent upon the chosen atom species, Rabi frequencies and Rydberg states, but to give an indication, ~\cite{Muller2009-uj} calculated that EIT can be used to perform the operation $\ket{+000} \mapsto 1/\sqrt{2} ( \ket{0000} + \ket{1111})$ with a fidelity in excess of 97\% with $^{87}$Rb. Higher fidelities may be achieved by an appropriate choice of the laser parameters and  Rydberg states, as discussed in~\cite{MacCormick2016-gc} and~\cite{Mansell2014-zw}, where the gating parameters were optimized for different spatial arrangements of the target qubits. Fidelities alone don't provide details of the underlying error channels so cannot be mapped to error correction thresholds --- leakage errors, for example, can be less harmful than Pauli errors~\cite{Suchara2015-ll}.

\begin{figure}[h]
    \centering
    \includegraphics[width=0.5\columnwidth]{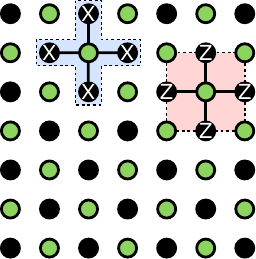}
    \caption{Arrangement of Rydberg atoms for a planar code. Green shaded qubits denote ancilla syndrome qubits used to measure stabilizer generators, and solid black qubits denote data qubits of the planar code. Using different species of atoms for syndrome and data qubits may help to reduce crosstalk during measurement~\cite{Beterov2015-zv}.}
    \label{fig:atom_arrangement}
\end{figure}

Our proposal requires atoms to be trapped in a lattice configuration like that shown in \figref{fig:atom_arrangement}. Deterministic loading of traps remains a major hurdle for Rydberg atom quantum computation, but methods to overcome this have been suggested, including starting with a partially loaded lattice and rearranging the qubits~\cite{Weiss2004-eg} --- this approach has been successfully used to construct 2D lattice geometries of $\sim 50$ qubits~\cite{Barredo2016-bf} with atomic separations of a few $\mu m$ using optical tweezers, which would be sufficient for a prototype device. It is not necessary to construct a perfect lattice, as low rates of missing qubits can be handled with no additional quantum processing~\cite{Auger2017-di}.

Once the atoms are trapped, error correction proceeds by repeatedly measuring the stabilizer generators of the surface code. For example, to measure a star stabilizer generator, the syndrome qubit is prepared in the $\ket{+}$ state, and then the EIT gate method is used to apply simultaneous CNOT gates controlled by the associated syndrome qubit, with the four surrounding data qubits as targets. A Hadamard gate is then applied to the syndrome qubit before it is measured in the computational basis. This process is shown in \figref{fig:star_measurement}. Measurement of plaquette stabilizer generators is performed in the same manner, but with Hadamard gates applied to each data qubit before and after the CNOT gates.

\begin{figure}[h]
    \centering
    \begin{subfigure}{0.45\linewidth}
        \includegraphics[width=\textwidth]{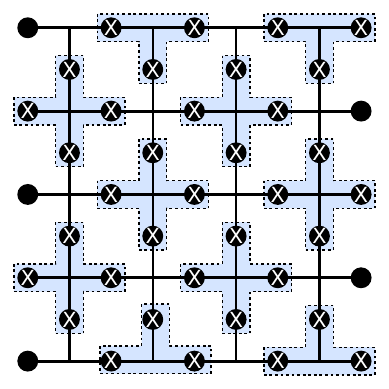}
        \caption{\label{fig:order_of_measurementa}}
    \end{subfigure}
    \hfill
    \begin{subfigure}{0.45\linewidth}
        \includegraphics[width=\textwidth]{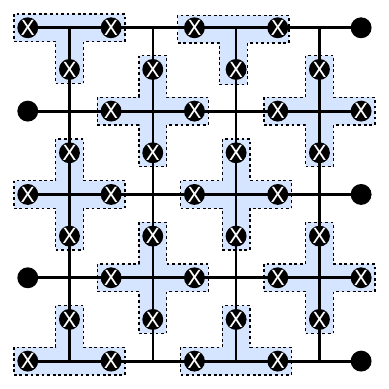}
        \caption{\label{fig:order_of_measurementb}}
    \end{subfigure}

    \begin{subfigure}{0.45\linewidth}
        \includegraphics[width=\textwidth]{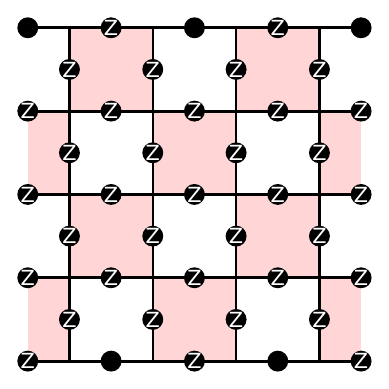}
        \caption{\label{fig:order_of_measurementc}}
    \end{subfigure}
    \hfill
    \begin{subfigure}{0.45\linewidth}
        \includegraphics[width=\textwidth]{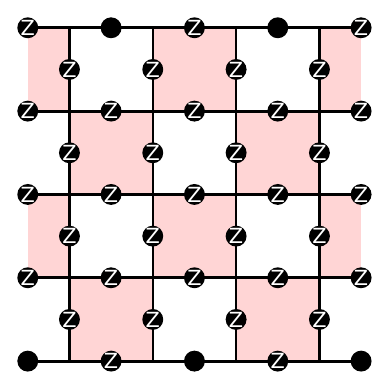}
        \caption{\label{fig:order_of_measurementd}}
    \end{subfigure}
    \caption{Stabilizer generators must be measured in at least four stages to ensure that each data qubit is only involved in a single interaction at any time. Each of the subfigures represents one stage of measurement, with all four stages required for one complete round of syndrome measurement. (\subref{fig:order_of_measurementa}) and (\subref{fig:order_of_measurementb}) show measurement of star stabilizer generators, and (\subref{fig:order_of_measurementc}) and (\subref{fig:order_of_measurementd}) show measurement of plaquette stabilizer generators.
    }
    \label{fig:order_of_measurement}
\end{figure}

Each data qubit can only be involved in one interaction at a given time, so the star and plaquette measurement operations must be performed in at least four separate stages, as shown in \figref{fig:order_of_measurement}. It should be noted that because of the scaling of the Van der Waals interaction with distance, the number of staggered measurements may need to be increased to avoid crosstalk between control and target qubits belonging to different stars and plaquettes. This staggered measurement pattern should not impact the overall speed of the computation significantly, as the readout stage is several orders of magnitude slower than the interaction stage, and the actual readout from the syndrome qubits can be performed simultaneously. Should measurement speed be increased, then the choice of measurement pattern will depend upon the trade-off between errors accumulating due to the delay between stabilizer generator measurements and errors occurring due to crosstalk during the EIT gates.

Fast, high-fidelity measurement is another outstanding challenge for Rydberg atom devices. Quantum nondemolition measurements with arrays of qubits have only been performed using relatively noisy electron-multiplying CCDs~\cite{Alberti2016-su} rather than discrete photon detectors. Such measurements take around 20 ms~\cite{Martinez_Dorantes2016-ik}, so this is currently the limiting factor for the clock speed of our scheme and will limit the computation speed to Hz frequencies until improvements are made. Crosstalk during measurement poses an additional problem, although suggestions for reducing crosstalk by using a two-species architecture~\cite{Beterov2015-zv} would be ideally suited to a surface code quantum computer, where rubidium atoms could be used for the frequently-measured syndrome qubits and caesium atoms could be used for the data qubits.

\begin{figure}[h]
    \centering
    \includegraphics[width=0.5\columnwidth]{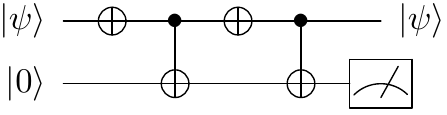}
    \caption{\label{fig:leakage_detection}Leakage detection circuit from~\cite{Preskill1997-tv}. If the top qubit is in the computational basis, then the bottom qubit (an ancilla) will be observed in the $\ket{1}$ state. If the top qubit has leaked or been lost, the bottom qubit will be observed in the $\ket{0}$ state. The top qubit can be reinitialized if leakage or loss occurs.}
\end{figure}

As atoms are non-binary systems and we are making extensive use of non-qubit Rydberg states of atoms, it is prudent to include some form of leakage and loss detection and reduction; the leakage detection circuit in \figref{fig:leakage_detection} can be used periodically for such a purpose~\cite{Preskill1997-tv}. Methods for dealing with leakage in the surface code were considered in~\cite{Suchara2015-ll}, which showed that low levels of leakage can be tolerated at the cost of a slightly lower error correction threshold. When a qubit has leaked or been lost, this qubit can be reset to a known state, e.g. $\ket{0}$, and the error correction can proceed as normal, with the decoder taking account of the increased probability of an error occurring on the leaked qubit. The frequency with which leakage detection needs to be performed will depend on the rate at which leakage errors occur; leakage detection will introduce additional errors so will ideally be performed as infrequently as possible.

In addition to syndrome measurement, performing full quantum computation by braiding defects or lattice surgery will require the ability to measure individual data qubits occasionally. This could be achieved by using CNOT gates and measuring syndrome qubits, therefore removing the requirement to be able to directly measure the data qubits.

\section{\label{sec:error_thresholds}Error thresholds}

We have performed a simulation of this scheme to obtain an error correction threshold --- the threshold is the critical physical qubit error rate below which increasing the number of qubits in the code reduces the logical error rate, meaning arbitrary quantum computations can be performed, providing there are enough qubits. The simulation uses the planar code with stabilizer generator measurements, as outlined in Sec.~\ref{sec:scheme}, in the presence of errors.

The error model used in the simulation is based around a single error parameter $p$ as follows. State preparation and measurement are assumed to result in the preparation or detection of orthogonal states respectively with probability $p$. Each multi-qubit EIT gate is modeled to act perfectly followed by depolarizing noise with probability $p$, i.e. for an $n$ qubit gate, each of the possible $4^n -1$ non-identity Pauli operations will occur with probability $p/(4^n -1)$. Single qubit gates, such as identity gates and Hadamard gates, are assumed to be free from error on the basis that such operations will generally have much lower error rates than other operations.

As mentioned in section \ref{sec:surface_codes}, measuring the surface code stabilizer generators leads to a discretization of errors: a more general error will collapse into a combination of Pauli errors~\cite{Nielsen2000-nv}. It is non-trivial to perform a simulation with physical errors or to directly relate physical errors to a simulatable error model. This error model therefore has been chosen in lieu of knowledge of the exact error channel and associated Pauli error rates, as in standard when obtaining quantum error correction thresholds; this allows for a comparison with thresholds obtained for other approaches, such as those in~\cite{Stephens2014-qc}.

Leakage errors, such as atom loss or excitation of unintended energy levels, were not considered in the simulation --- such a simulation is left for future work.

\begin{figure}[h]
    \centering
    \includegraphics[width=\columnwidth]{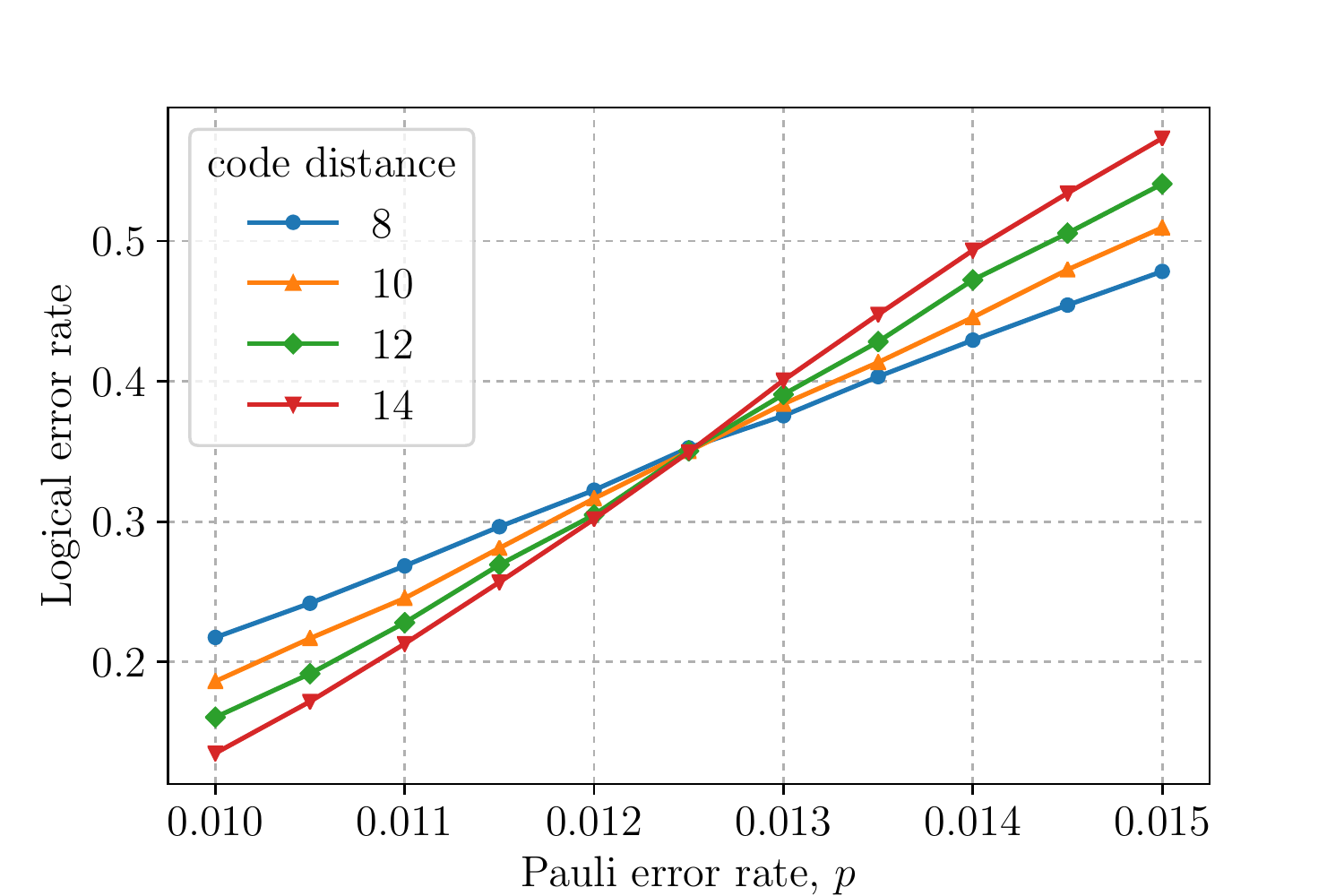}
    \caption{Logical error rates from error simulations for code distances 8, 10, 12 and 14. The crossing point gives the resulting error threshold of 1.25\%.}
    \label{fig:threshold}
\end{figure}

Planar codes with code distance $d=8,10,12$ and $14$ were simulated for $2d$ rounds of syndrome measurement using the scheme outlined in Sec.~\ref{sec:scheme} and the above error model, and a minimum weight perfect matching algorithm was used for decoding. \figref{fig:threshold} shows the logical error rates obtained during the simulations. The error threshold is given by the crossing point in this plot~\cite{Dennis2002-de, Wang2003-mh} resulting in a threshold of $p_{th}\approx 1.25\%$ for our chosen error model.

\subsection{\label{sec:correlated_errors}Correlated errors}

The Van der Waals interaction between atoms scales with $1/{R^6}$, where $R$ is the separation between atoms. This polynomial decay means that there may be a non-negligible crosstalk between distant qubits during multi-qubit gates, which could cause correlated errors between non-neighboring qubits. Similar errors were considered for the surface code in \cite{Fowler2014-en}, which found that logical error suppression could be achieved even in the extreme case of quadratically decaying interactions.

\section{\label{sec:summary}Summary}

We have proposed a new scheme for fault-tolerant quantum computation with Rydberg atoms. Our proposal uses EIT to perform multi-qubit gates for syndrome extraction, and we suggest methods to mitigate the effects of leakage and qubit loss. We have found a threshold of 1.25\% for an error model based on this scheme, which we hope will provide an initial target for experimentalists looking to build a prototype quantum computer with Rydberg atoms. Prospects for initial scalability are good, with arrays with on the order of $10^4$ atoms being realistically achievable~\cite{Saffman2016-je}. Larger numbers of qubits would be desirable in the long run, but this should be satisfactory for early devices attempting to demonstrate quantum speedup.

Experimentally achieving quantum operations with sufficiently high fidelities and low loss rates remains a challenge with Rydberg atoms, but this is mostly due to engineering obstacles rather than physical limitations, so we are optimistic that large improvements will be made. While quantum gates can be performed at MHz frequencies, slow measurement for arrays of atoms currently limits the potential clock speed of our scheme to the order of a few tens of Hz, so this is a key area for improvement. A thorough error analysis of multi-qubit gates based around EIT is required to determine whether sufficiently low error rates can be achieved --- it is likely that improvements will be needed to achieve the error rates below 1.25\% required for reliable quantum computation. If necessary, the EIT gates in our proposal can easily be replaced by another multi-qubit interaction without significantly affecting the rest of the scheme.

It should be noted that a threshold alone cannot be used to verify that a scheme will work for surface code based quantum computation --- a more convincing analysis is to experimentally demonstrate that a larger system has superior error suppression compared to a smaller system, as has been accomplished with bit-flip errors on superconducting qubits~\cite{Kelly2015-gp}.

Our findings suggest that while there are many advantageous features of Rydberg atoms, gate fidelities need to be improved before fault-tolerant universal quantum computation can be achieved --- experiments based on other implementations of fault-tolerant quantum computation, such as superconducting qubits~\cite{Barends2014-tf} and trapped ions~\cite{Ballance2016-hu} are currently ahead of Rydberg atoms. We nonetheless believe Rydberg atoms are a candidate for building a fault-tolerant quantum computer.

\begin{acknowledgments}

JMA is funded by EPSRC. The authors acknowledge the use of the UCL Legion High Performance Computing Facility (Legion@UCL), and associated support services, in the completion of this work. Blossom V~\cite{Kolmogorov2009-md} was used for minimum weight perfect matching in simulations.

\end{acknowledgments}


\begin{thebibliography}{42}%
\makeatletter
\providecommand \@ifxundefined [1]{%
 \@ifx{#1\undefined}
}%
\providecommand \@ifnum [1]{%
 \ifnum #1\expandafter \@firstoftwo
 \else \expandafter \@secondoftwo
 \fi
}%
\providecommand \@ifx [1]{%
 \ifx #1\expandafter \@firstoftwo
 \else \expandafter \@secondoftwo
 \fi
}%
\providecommand \natexlab [1]{#1}%
\providecommand \enquote  [1]{``#1''}%
\providecommand \bibnamefont  [1]{#1}%
\providecommand \bibfnamefont [1]{#1}%
\providecommand \citenamefont [1]{#1}%
\providecommand \href@noop [0]{\@secondoftwo}%
\providecommand \href [0]{\begingroup \@sanitize@url \@href}%
\providecommand \@href[1]{\@@startlink{#1}\@@href}%
\providecommand \@@href[1]{\endgroup#1\@@endlink}%
\providecommand \@sanitize@url [0]{\catcode `\\12\catcode `\$12\catcode
  `\&12\catcode `\#12\catcode `\^12\catcode `\_12\catcode `\%12\relax}%
\providecommand \@@startlink[1]{}%
\providecommand \@@endlink[0]{}%
\providecommand \url  [0]{\begingroup\@sanitize@url \@url }%
\providecommand \@url [1]{\endgroup\@href {#1}{\urlprefix }}%
\providecommand \urlprefix  [0]{URL }%
\providecommand \Eprint [0]{\href }%
\providecommand \doibase [0]{http://dx.doi.org/}%
\providecommand \selectlanguage [0]{\@gobble}%
\providecommand \bibinfo  [0]{\@secondoftwo}%
\providecommand \bibfield  [0]{\@secondoftwo}%
\providecommand \translation [1]{[#1]}%
\providecommand \BibitemOpen [0]{}%
\providecommand \bibitemStop [0]{}%
\providecommand \bibitemNoStop [0]{.\EOS\space}%
\providecommand \EOS [0]{\spacefactor3000\relax}%
\providecommand \BibitemShut  [1]{\csname bibitem#1\endcsname}%
\let\auto@bib@innerbib\@empty
\bibitem [{\citenamefont {Saffman}\ \emph {et~al.}(2010)\citenamefont
  {Saffman}, \citenamefont {Walker},\ and\ \citenamefont
  {M{\o}lmer}}]{Saffman2010-ea}%
  \BibitemOpen
  \bibfield  {author} {\bibinfo {author} {\bibfnamefont {M.}~\bibnamefont
  {Saffman}}, \bibinfo {author} {\bibfnamefont {T.~G.}\ \bibnamefont {Walker}},
  \ and\ \bibinfo {author} {\bibfnamefont {K.}~\bibnamefont {M{\o}lmer}},\
  }\href {\doibase 10.1103/RevModPhys.82.2313} {\bibfield  {journal} {\bibinfo
  {journal} {Rev. Mod. Phys.}\ }\textbf {\bibinfo {volume} {82}},\ \bibinfo
  {pages} {2313} (\bibinfo {year} {2010})},\ \Eprint
  {http://arxiv.org/abs/0909.4777} {arXiv:0909.4777 [quant-ph]} \BibitemShut
  {NoStop}%
\bibitem [{\citenamefont {Bernien}\ \emph {et~al.}(2017)\citenamefont
  {Bernien}, \citenamefont {Schwartz}, \citenamefont {Keesling}, \citenamefont
  {Levine}, \citenamefont {Omran}, \citenamefont {Pichler}, \citenamefont
  {Choi}, \citenamefont {Zibrov}, \citenamefont {Endres}, \citenamefont
  {Greiner}, \citenamefont {Vuleti{\'c}},\ and\ \citenamefont
  {Lukin}}]{Bernien2017-hn}%
  \BibitemOpen
  \bibfield  {author} {\bibinfo {author} {\bibfnamefont {H.}~\bibnamefont
  {Bernien}}, \bibinfo {author} {\bibfnamefont {S.}~\bibnamefont {Schwartz}},
  \bibinfo {author} {\bibfnamefont {A.}~\bibnamefont {Keesling}}, \bibinfo
  {author} {\bibfnamefont {H.}~\bibnamefont {Levine}}, \bibinfo {author}
  {\bibfnamefont {A.}~\bibnamefont {Omran}}, \bibinfo {author} {\bibfnamefont
  {H.}~\bibnamefont {Pichler}}, \bibinfo {author} {\bibfnamefont
  {S.}~\bibnamefont {Choi}}, \bibinfo {author} {\bibfnamefont {A.~S.}\
  \bibnamefont {Zibrov}}, \bibinfo {author} {\bibfnamefont {M.}~\bibnamefont
  {Endres}}, \bibinfo {author} {\bibfnamefont {M.}~\bibnamefont {Greiner}},
  \bibinfo {author} {\bibfnamefont {V.}~\bibnamefont {Vuleti{\'c}}}, \ and\
  \bibinfo {author} {\bibfnamefont {M.~D.}\ \bibnamefont {Lukin}},\ }\href@noop
  {} {\  (\bibinfo {year} {2017})},\ \Eprint {http://arxiv.org/abs/1707.04344}
  {arXiv:1707.04344 [quant-ph]} \BibitemShut {NoStop}%
\bibitem [{\citenamefont {Brion}\ \emph {et~al.}(2008)\citenamefont {Brion},
  \citenamefont {Pedersen}, \citenamefont {Saffman},\ and\ \citenamefont
  {M{\o}lmer}}]{Brion2008-iv}%
  \BibitemOpen
  \bibfield  {author} {\bibinfo {author} {\bibfnamefont {E.}~\bibnamefont
  {Brion}}, \bibinfo {author} {\bibfnamefont {L.~H.}\ \bibnamefont {Pedersen}},
  \bibinfo {author} {\bibfnamefont {M.}~\bibnamefont {Saffman}}, \ and\
  \bibinfo {author} {\bibfnamefont {K.}~\bibnamefont {M{\o}lmer}},\ }\href
  {\doibase 10.1103/PhysRevLett.100.110506} {\bibfield  {journal} {\bibinfo
  {journal} {Phys. Rev. Lett.}\ }\textbf {\bibinfo {volume} {100}},\ \bibinfo
  {pages} {110506} (\bibinfo {year} {2008})},\ \Eprint
  {http://arxiv.org/abs/0710.1717} {arXiv:0710.1717 [quant-ph]} \BibitemShut
  {NoStop}%
\bibitem [{\citenamefont {Crow}\ \emph {et~al.}(2016)\citenamefont {Crow},
  \citenamefont {Joynt},\ and\ \citenamefont {Saffman}}]{Crow2016-ay}%
  \BibitemOpen
  \bibfield  {author} {\bibinfo {author} {\bibfnamefont {D.}~\bibnamefont
  {Crow}}, \bibinfo {author} {\bibfnamefont {R.}~\bibnamefont {Joynt}}, \ and\
  \bibinfo {author} {\bibfnamefont {M.}~\bibnamefont {Saffman}},\ }\href
  {\doibase 10.1103/PhysRevLett.117.130503} {\bibfield  {journal} {\bibinfo
  {journal} {Phys. Rev. Lett.}\ }\textbf {\bibinfo {volume} {117}},\ \bibinfo
  {pages} {130503} (\bibinfo {year} {2016})},\ \Eprint
  {http://arxiv.org/abs/1510.08359} {arXiv:1510.08359 [quant-ph]} \BibitemShut
  {NoStop}%
\bibitem [{\citenamefont {Isenhower}\ \emph {et~al.}(2011)\citenamefont
  {Isenhower}, \citenamefont {Saffman},\ and\ \citenamefont
  {M{\o}lmer}}]{Isenhower2011-wh}%
  \BibitemOpen
  \bibfield  {author} {\bibinfo {author} {\bibfnamefont {L.}~\bibnamefont
  {Isenhower}}, \bibinfo {author} {\bibfnamefont {M.}~\bibnamefont {Saffman}},
  \ and\ \bibinfo {author} {\bibfnamefont {K.}~\bibnamefont {M{\o}lmer}},\
  }\href {\doibase 10.1007/s11128-011-0292-4} {\bibfield  {journal} {\bibinfo
  {journal} {Quantum Inf. Process.}\ }\textbf {\bibinfo {volume} {10}},\
  \bibinfo {pages} {755} (\bibinfo {year} {2011})},\ \Eprint
  {http://arxiv.org/abs/1104.3916} {arXiv:1104.3916 [quant-ph]} \BibitemShut
  {NoStop}%
\bibitem [{\citenamefont {Brion}\ \emph {et~al.}(2007)\citenamefont {Brion},
  \citenamefont {Pedersen}, \citenamefont {M{\o}lmer}, \citenamefont {Chutia},\
  and\ \citenamefont {Saffman}}]{Brion2007-ly}%
  \BibitemOpen
  \bibfield  {author} {\bibinfo {author} {\bibfnamefont {E.}~\bibnamefont
  {Brion}}, \bibinfo {author} {\bibfnamefont {L.~H.}\ \bibnamefont {Pedersen}},
  \bibinfo {author} {\bibfnamefont {K.}~\bibnamefont {M{\o}lmer}}, \bibinfo
  {author} {\bibfnamefont {S.}~\bibnamefont {Chutia}}, \ and\ \bibinfo {author}
  {\bibfnamefont {M.}~\bibnamefont {Saffman}},\ }\href {\doibase
  10.1103/PhysRevA.75.032328} {\bibfield  {journal} {\bibinfo  {journal} {Phys.
  Rev. A}\ }\textbf {\bibinfo {volume} {75}},\ \bibinfo {pages} {032328}
  (\bibinfo {year} {2007})},\ \Eprint {http://arxiv.org/abs/quant-ph/0611246}
  {arXiv:quant-ph/0611246} \BibitemShut {NoStop}%
\bibitem [{\citenamefont {Kitaev}(2003)}]{Kitaev2003-bz}%
  \BibitemOpen
  \bibfield  {author} {\bibinfo {author} {\bibfnamefont {A.~Y.}\ \bibnamefont
  {Kitaev}},\ }\href {\doibase 10.1016/S0003-4916(02)00018-0} {\bibfield
  {journal} {\bibinfo  {journal} {Ann. Phys.}\ }\textbf {\bibinfo {volume}
  {303}},\ \bibinfo {pages} {2} (\bibinfo {year} {2003})},\ \Eprint
  {http://arxiv.org/abs/quant-ph/9707021} {arXiv:quant-ph/9707021} \BibitemShut
  {NoStop}%
\bibitem [{\citenamefont {Bravyi}\ and\ \citenamefont
  {Yu.~Kitaev}(1998)}]{Bravyi1998-kf}%
  \BibitemOpen
  \bibfield  {author} {\bibinfo {author} {\bibfnamefont {S.~B.}\ \bibnamefont
  {Bravyi}}\ and\ \bibinfo {author} {\bibfnamefont {A.}~\bibnamefont
  {Yu.~Kitaev}},\ }\href@noop {} {\  (\bibinfo {year} {1998})},\ \Eprint
  {http://arxiv.org/abs/quant-ph/9811052} {arXiv:quant-ph/9811052} \BibitemShut
  {NoStop}%
\bibitem [{\citenamefont {Bergamini}\ \emph {et~al.}(2004)\citenamefont
  {Bergamini}, \citenamefont {Darqui{\'e}}, \citenamefont {Jones},
  \citenamefont {Jacubowiez}, \citenamefont {Browaeys},\ and\ \citenamefont
  {Grangier}}]{Bergamini2004-bf}%
  \BibitemOpen
  \bibfield  {author} {\bibinfo {author} {\bibfnamefont {S.}~\bibnamefont
  {Bergamini}}, \bibinfo {author} {\bibfnamefont {B.}~\bibnamefont
  {Darqui{\'e}}}, \bibinfo {author} {\bibfnamefont {M.}~\bibnamefont {Jones}},
  \bibinfo {author} {\bibfnamefont {L.}~\bibnamefont {Jacubowiez}}, \bibinfo
  {author} {\bibfnamefont {A.}~\bibnamefont {Browaeys}}, \ and\ \bibinfo
  {author} {\bibfnamefont {P.}~\bibnamefont {Grangier}},\ }\href {\doibase
  10.1364/JOSAB.21.001889} {\bibfield  {journal} {\bibinfo  {journal} {J. Opt.
  Soc. Am. B, JOSAB}\ }\textbf {\bibinfo {volume} {21}},\ \bibinfo {pages}
  {1889} (\bibinfo {year} {2004})}\BibitemShut {NoStop}%
\bibitem [{\citenamefont {Weimer}\ \emph {et~al.}(2010)\citenamefont {Weimer},
  \citenamefont {M{\"u}ller}, \citenamefont {Lesanovsky}, \citenamefont
  {Zoller},\ and\ \citenamefont {B{\"u}chler}}]{Weimer2010-hf}%
  \BibitemOpen
  \bibfield  {author} {\bibinfo {author} {\bibfnamefont {H.}~\bibnamefont
  {Weimer}}, \bibinfo {author} {\bibfnamefont {M.}~\bibnamefont {M{\"u}ller}},
  \bibinfo {author} {\bibfnamefont {I.}~\bibnamefont {Lesanovsky}}, \bibinfo
  {author} {\bibfnamefont {P.}~\bibnamefont {Zoller}}, \ and\ \bibinfo {author}
  {\bibfnamefont {H.~P.}\ \bibnamefont {B{\"u}chler}},\ }\href {\doibase
  10.1038/nphys1614} {\bibfield  {journal} {\bibinfo  {journal} {Nat. Phys.}\
  }\textbf {\bibinfo {volume} {6}},\ \bibinfo {pages} {382} (\bibinfo {year}
  {2010})},\ \Eprint {http://arxiv.org/abs/0907.1657} {arXiv:0907.1657
  [quant-ph]} \BibitemShut {NoStop}%
\bibitem [{\citenamefont {Weimer}\ \emph {et~al.}(2011)\citenamefont {Weimer},
  \citenamefont {M{\"u}ller}, \citenamefont {B{\"u}chler},\ and\ \citenamefont
  {Lesanovsky}}]{Weimer2011-hr}%
  \BibitemOpen
  \bibfield  {author} {\bibinfo {author} {\bibfnamefont {H.}~\bibnamefont
  {Weimer}}, \bibinfo {author} {\bibfnamefont {M.}~\bibnamefont {M{\"u}ller}},
  \bibinfo {author} {\bibfnamefont {H.~P.}\ \bibnamefont {B{\"u}chler}}, \ and\
  \bibinfo {author} {\bibfnamefont {I.}~\bibnamefont {Lesanovsky}},\ }\href
  {\doibase 10.1007/s11128-011-0303-5} {\bibfield  {journal} {\bibinfo
  {journal} {Quantum Inf. Process.}\ }\textbf {\bibinfo {volume} {10}},\
  \bibinfo {pages} {885} (\bibinfo {year} {2011})},\ \Eprint
  {http://arxiv.org/abs/1104.3081} {arXiv:1104.3081 [quant-ph]} \BibitemShut
  {NoStop}%
\bibitem [{\citenamefont {Xia}\ \emph {et~al.}(2015)\citenamefont {Xia},
  \citenamefont {Lichtman}, \citenamefont {Maller}, \citenamefont {Carr},
  \citenamefont {Piotrowicz}, \citenamefont {Isenhower},\ and\ \citenamefont
  {Saffman}}]{Xia2015-dm}%
  \BibitemOpen
  \bibfield  {author} {\bibinfo {author} {\bibfnamefont {T.}~\bibnamefont
  {Xia}}, \bibinfo {author} {\bibfnamefont {M.}~\bibnamefont {Lichtman}},
  \bibinfo {author} {\bibfnamefont {K.}~\bibnamefont {Maller}}, \bibinfo
  {author} {\bibfnamefont {A.~W.}\ \bibnamefont {Carr}}, \bibinfo {author}
  {\bibfnamefont {M.~J.}\ \bibnamefont {Piotrowicz}}, \bibinfo {author}
  {\bibfnamefont {L.}~\bibnamefont {Isenhower}}, \ and\ \bibinfo {author}
  {\bibfnamefont {M.}~\bibnamefont {Saffman}},\ }\href {\doibase
  10.1103/PhysRevLett.114.100503} {\bibfield  {journal} {\bibinfo  {journal}
  {Phys. Rev. Lett.}\ }\textbf {\bibinfo {volume} {114}},\ \bibinfo {pages}
  {100503} (\bibinfo {year} {2015})},\ \Eprint
  {http://arxiv.org/abs/1501.02041} {arXiv:1501.02041 [quant-ph]} \BibitemShut
  {NoStop}%
\bibitem [{\citenamefont {Wang}\ \emph {et~al.}(2016)\citenamefont {Wang},
  \citenamefont {Kumar}, \citenamefont {Wu},\ and\ \citenamefont
  {Weiss}}]{Wang2016-tq}%
  \BibitemOpen
  \bibfield  {author} {\bibinfo {author} {\bibfnamefont {Y.}~\bibnamefont
  {Wang}}, \bibinfo {author} {\bibfnamefont {A.}~\bibnamefont {Kumar}},
  \bibinfo {author} {\bibfnamefont {T.-Y.}\ \bibnamefont {Wu}}, \ and\ \bibinfo
  {author} {\bibfnamefont {D.~S.}\ \bibnamefont {Weiss}},\ }\href {\doibase
  10.1126/science.aaf2581} {\bibfield  {journal} {\bibinfo  {journal}
  {Science}\ }\textbf {\bibinfo {volume} {352}},\ \bibinfo {pages} {1562}
  (\bibinfo {year} {2016})},\ \Eprint {http://arxiv.org/abs/1601.03639}
  {arXiv:1601.03639 [quant-ph]} \BibitemShut {NoStop}%
\bibitem [{\citenamefont {Jau}\ \emph {et~al.}(2015)\citenamefont {Jau},
  \citenamefont {Hankin}, \citenamefont {Keating}, \citenamefont {Deutsch},\
  and\ \citenamefont {Biedermann}}]{Jau2015-tf}%
  \BibitemOpen
  \bibfield  {author} {\bibinfo {author} {\bibfnamefont {Y.-Y.}\ \bibnamefont
  {Jau}}, \bibinfo {author} {\bibfnamefont {A.~M.}\ \bibnamefont {Hankin}},
  \bibinfo {author} {\bibfnamefont {T.}~\bibnamefont {Keating}}, \bibinfo
  {author} {\bibfnamefont {I.~H.}\ \bibnamefont {Deutsch}}, \ and\ \bibinfo
  {author} {\bibfnamefont {G.~W.}\ \bibnamefont {Biedermann}},\ }\href
  {\doibase 10.1038/nphys3487} {\bibfield  {journal} {\bibinfo  {journal} {Nat.
  Phys.}\ }\textbf {\bibinfo {volume} {12}},\ \bibinfo {pages} {71} (\bibinfo
  {year} {2015})},\ \Eprint {http://arxiv.org/abs/1501.03862} {arXiv:1501.03862
  [quant-ph]} \BibitemShut {NoStop}%
\bibitem [{\citenamefont {Maller}\ \emph {et~al.}(2015)\citenamefont {Maller},
  \citenamefont {Lichtman}, \citenamefont {Xia}, \citenamefont {Sun},
  \citenamefont {Piotrowicz}, \citenamefont {Carr}, \citenamefont {Isenhower},\
  and\ \citenamefont {Saffman}}]{Maller2015-ye}%
  \BibitemOpen
  \bibfield  {author} {\bibinfo {author} {\bibfnamefont {K.~M.}\ \bibnamefont
  {Maller}}, \bibinfo {author} {\bibfnamefont {M.~T.}\ \bibnamefont
  {Lichtman}}, \bibinfo {author} {\bibfnamefont {T.}~\bibnamefont {Xia}},
  \bibinfo {author} {\bibfnamefont {Y.}~\bibnamefont {Sun}}, \bibinfo {author}
  {\bibfnamefont {M.~J.}\ \bibnamefont {Piotrowicz}}, \bibinfo {author}
  {\bibfnamefont {A.~W.}\ \bibnamefont {Carr}}, \bibinfo {author}
  {\bibfnamefont {L.}~\bibnamefont {Isenhower}}, \ and\ \bibinfo {author}
  {\bibfnamefont {M.}~\bibnamefont {Saffman}},\ }\href {\doibase
  10.1103/PhysRevA.92.022336} {\bibfield  {journal} {\bibinfo  {journal} {Phys.
  Rev. A}\ }\textbf {\bibinfo {volume} {92}},\ \bibinfo {pages} {022336}
  (\bibinfo {year} {2015})},\ \Eprint {http://arxiv.org/abs/1506.06416}
  {arXiv:1506.06416 [quant-ph]} \BibitemShut {NoStop}%
\bibitem [{\citenamefont {Saffman}(2016)}]{Saffman2016-je}%
  \BibitemOpen
  \bibfield  {author} {\bibinfo {author} {\bibfnamefont {M.}~\bibnamefont
  {Saffman}},\ }\href {\doibase 10.1088/0953-4075/49/20/202001} {\bibfield
  {journal} {\bibinfo  {journal} {J. Phys. B At. Mol. Opt. Phys.}\ }\textbf
  {\bibinfo {volume} {49}},\ \bibinfo {pages} {202001} (\bibinfo {year}
  {2016})},\ \Eprint {http://arxiv.org/abs/1605.05207} {arXiv:1605.05207
  [quant-ph]} \BibitemShut {NoStop}%
\bibitem [{\citenamefont {Horsman}\ \emph {et~al.}(2012)\citenamefont
  {Horsman}, \citenamefont {Fowler}, \citenamefont {Devitt},\ and\
  \citenamefont {Van~Meter}}]{Horsman2012-wh}%
  \BibitemOpen
  \bibfield  {author} {\bibinfo {author} {\bibfnamefont {C.}~\bibnamefont
  {Horsman}}, \bibinfo {author} {\bibfnamefont {A.~G.}\ \bibnamefont {Fowler}},
  \bibinfo {author} {\bibfnamefont {S.}~\bibnamefont {Devitt}}, \ and\ \bibinfo
  {author} {\bibfnamefont {R.}~\bibnamefont {Van~Meter}},\ }\href {\doibase
  10.1088/1367-2630/14/12/123011} {\bibfield  {journal} {\bibinfo  {journal}
  {New J. Phys.}\ }\textbf {\bibinfo {volume} {14}},\ \bibinfo {pages} {123011}
  (\bibinfo {year} {2012})},\ \Eprint {http://arxiv.org/abs/1111.4022}
  {arXiv:1111.4022 [quant-ph]} \BibitemShut {NoStop}%
\bibitem [{\citenamefont {Fowler}\ \emph {et~al.}(2012)\citenamefont {Fowler},
  \citenamefont {Mariantoni}, \citenamefont {Martinis},\ and\ \citenamefont
  {Cleland}}]{Fowler2012-yi}%
  \BibitemOpen
  \bibfield  {author} {\bibinfo {author} {\bibfnamefont {A.~G.}\ \bibnamefont
  {Fowler}}, \bibinfo {author} {\bibfnamefont {M.}~\bibnamefont {Mariantoni}},
  \bibinfo {author} {\bibfnamefont {J.~M.}\ \bibnamefont {Martinis}}, \ and\
  \bibinfo {author} {\bibfnamefont {A.~N.}\ \bibnamefont {Cleland}},\ }\href
  {\doibase 10.1103/PhysRevA.86.032324} {\bibfield  {journal} {\bibinfo
  {journal} {Phys. Rev. A}\ }\textbf {\bibinfo {volume} {86}},\ \bibinfo
  {pages} {032324} (\bibinfo {year} {2012})},\ \Eprint
  {http://arxiv.org/abs/1208.0928} {arXiv:1208.0928 [quant-ph]} \BibitemShut
  {NoStop}%
\bibitem [{\citenamefont {Nielsen}\ and\ \citenamefont
  {Chuang}(2000)}]{Nielsen2000-nv}%
  \BibitemOpen
  \bibfield  {author} {\bibinfo {author} {\bibfnamefont {M.~A.}\ \bibnamefont
  {Nielsen}}\ and\ \bibinfo {author} {\bibfnamefont {I.~L.}\ \bibnamefont
  {Chuang}},\ }\href@noop {} {{\selectlanguage {en}\emph {\bibinfo {title}
  {Quantum Computation and Quantum Information}}}}\ (\bibinfo  {publisher}
  {Cambridge University Press},\ \bibinfo {year} {2000})\BibitemShut {NoStop}%
\bibitem [{\citenamefont {Dennis}\ \emph {et~al.}(2002)\citenamefont {Dennis},
  \citenamefont {Kitaev}, \citenamefont {Landahl},\ and\ \citenamefont
  {Preskill}}]{Dennis2002-de}%
  \BibitemOpen
  \bibfield  {author} {\bibinfo {author} {\bibfnamefont {E.}~\bibnamefont
  {Dennis}}, \bibinfo {author} {\bibfnamefont {A.}~\bibnamefont {Kitaev}},
  \bibinfo {author} {\bibfnamefont {A.}~\bibnamefont {Landahl}}, \ and\
  \bibinfo {author} {\bibfnamefont {J.}~\bibnamefont {Preskill}},\ }\href
  {\doibase 10.1063/1.1499754} {\bibfield  {journal} {\bibinfo  {journal} {J.
  Math. Phys.}\ }\textbf {\bibinfo {volume} {43}},\ \bibinfo {pages} {4452}
  (\bibinfo {year} {2002})},\ \Eprint {http://arxiv.org/abs/quant-ph/0110143}
  {arXiv:quant-ph/0110143} \BibitemShut {NoStop}%
\bibitem [{\citenamefont {Duclos-Cianci}\ and\ \citenamefont
  {Poulin}(2010)}]{Duclos-Cianci2010-qd}%
  \BibitemOpen
  \bibfield  {author} {\bibinfo {author} {\bibfnamefont {G.}~\bibnamefont
  {Duclos-Cianci}}\ and\ \bibinfo {author} {\bibfnamefont {D.}~\bibnamefont
  {Poulin}},\ }\href {\doibase 10.1103/PhysRevLett.104.050504} {\bibfield
  {journal} {\bibinfo  {journal} {Phys. Rev. Lett.}\ }\textbf {\bibinfo
  {volume} {104}},\ \bibinfo {pages} {050504} (\bibinfo {year} {2010})},\
  \Eprint {http://arxiv.org/abs/0911.0581} {arXiv:0911.0581 [quant-ph]}
  \BibitemShut {NoStop}%
\bibitem [{\citenamefont {Herold}\ \emph {et~al.}(2015)\citenamefont {Herold},
  \citenamefont {Campbell}, \citenamefont {Eisert},\ and\ \citenamefont
  {Kastoryano}}]{Herold2015-oc}%
  \BibitemOpen
  \bibfield  {author} {\bibinfo {author} {\bibfnamefont {M.}~\bibnamefont
  {Herold}}, \bibinfo {author} {\bibfnamefont {E.~T.}\ \bibnamefont
  {Campbell}}, \bibinfo {author} {\bibfnamefont {J.}~\bibnamefont {Eisert}}, \
  and\ \bibinfo {author} {\bibfnamefont {M.~J.}\ \bibnamefont {Kastoryano}},\
  }\href {\doibase 10.1038/npjqi.2015.10} {\bibfield  {journal} {\bibinfo
  {journal} {npj Quantum Information}\ }\textbf {\bibinfo {volume} {1}},\
  \bibinfo {pages} {15010} (\bibinfo {year} {2015})},\ \Eprint
  {http://arxiv.org/abs/1406.2338} {arXiv:1406.2338 [quant-ph]} \BibitemShut
  {NoStop}%
\bibitem [{\citenamefont {Bravyi}\ \emph {et~al.}(2014)\citenamefont {Bravyi},
  \citenamefont {Suchara},\ and\ \citenamefont {Vargo}}]{Bravyi2014-kg}%
  \BibitemOpen
  \bibfield  {author} {\bibinfo {author} {\bibfnamefont {S.}~\bibnamefont
  {Bravyi}}, \bibinfo {author} {\bibfnamefont {M.}~\bibnamefont {Suchara}}, \
  and\ \bibinfo {author} {\bibfnamefont {A.}~\bibnamefont {Vargo}},\ }\href
  {\doibase 10.1103/PhysRevA.90.032326} {\bibfield  {journal} {\bibinfo
  {journal} {Phys. Rev. A}\ }\textbf {\bibinfo {volume} {90}},\ \bibinfo
  {pages} {032326} (\bibinfo {year} {2014})},\ \Eprint
  {http://arxiv.org/abs/1405.4883} {arXiv:1405.4883 [quant-ph]} \BibitemShut
  {NoStop}%
\bibitem [{\citenamefont {Delfosse}\ and\ \citenamefont
  {Z{\'e}mor}(2017)}]{Delfosse2017-pa}%
  \BibitemOpen
  \bibfield  {author} {\bibinfo {author} {\bibfnamefont {N.}~\bibnamefont
  {Delfosse}}\ and\ \bibinfo {author} {\bibfnamefont {G.}~\bibnamefont
  {Z{\'e}mor}},\ }\href@noop {} {\  (\bibinfo {year} {2017})},\ \Eprint
  {http://arxiv.org/abs/1703.01517} {arXiv:1703.01517 [quant-ph]} \BibitemShut
  {NoStop}%
\bibitem [{\citenamefont {M{\"u}ller}\ \emph {et~al.}(2009)\citenamefont
  {M{\"u}ller}, \citenamefont {Lesanovsky}, \citenamefont {Weimer},
  \citenamefont {B{\"u}chler},\ and\ \citenamefont {Zoller}}]{Muller2009-uj}%
  \BibitemOpen
  \bibfield  {author} {\bibinfo {author} {\bibfnamefont {M.}~\bibnamefont
  {M{\"u}ller}}, \bibinfo {author} {\bibfnamefont {I.}~\bibnamefont
  {Lesanovsky}}, \bibinfo {author} {\bibfnamefont {H.}~\bibnamefont {Weimer}},
  \bibinfo {author} {\bibfnamefont {H.~P.}\ \bibnamefont {B{\"u}chler}}, \ and\
  \bibinfo {author} {\bibfnamefont {P.}~\bibnamefont {Zoller}},\ }\href
  {\doibase 10.1103/PhysRevLett.102.170502} {\bibfield  {journal} {\bibinfo
  {journal} {Phys. Rev. Lett.}\ }\textbf {\bibinfo {volume} {102}},\ \bibinfo
  {pages} {170502} (\bibinfo {year} {2009})},\ \Eprint
  {http://arxiv.org/abs/0811.1155} {arXiv:0811.1155 [quant-ph]} \BibitemShut
  {NoStop}%
\bibitem [{\citenamefont {MacCormick}\ \emph {et~al.}(2016)\citenamefont
  {MacCormick}, \citenamefont {Bergamini}, \citenamefont {Mansell},
  \citenamefont {Cable},\ and\ \citenamefont {Modi}}]{MacCormick2016-gc}%
  \BibitemOpen
  \bibfield  {author} {\bibinfo {author} {\bibfnamefont {C.}~\bibnamefont
  {MacCormick}}, \bibinfo {author} {\bibfnamefont {S.}~\bibnamefont
  {Bergamini}}, \bibinfo {author} {\bibfnamefont {C.}~\bibnamefont {Mansell}},
  \bibinfo {author} {\bibfnamefont {H.}~\bibnamefont {Cable}}, \ and\ \bibinfo
  {author} {\bibfnamefont {K.}~\bibnamefont {Modi}},\ }\href {\doibase
  10.1103/PhysRevA.93.023805} {\bibfield  {journal} {\bibinfo  {journal} {Phys.
  Rev. A}\ }\textbf {\bibinfo {volume} {93}},\ \bibinfo {pages} {023805}
  (\bibinfo {year} {2016})},\ \Eprint {http://arxiv.org/abs/1511.02741}
  {arXiv:1511.02741 [quant-ph]} \BibitemShut {NoStop}%
\bibitem [{\citenamefont {Mansell}\ and\ \citenamefont
  {Bergamini}(2014)}]{Mansell2014-zw}%
  \BibitemOpen
  \bibfield  {author} {\bibinfo {author} {\bibfnamefont {C.~W.}\ \bibnamefont
  {Mansell}}\ and\ \bibinfo {author} {\bibfnamefont {S.}~\bibnamefont
  {Bergamini}},\ }\href {\doibase 10.1088/1367-2630/16/5/053045} {\bibfield
  {journal} {\bibinfo  {journal} {New J. Phys.}\ }\textbf {\bibinfo {volume}
  {16}},\ \bibinfo {pages} {053045} (\bibinfo {year} {2014})},\ \Eprint
  {http://arxiv.org/abs/1309.7920} {arXiv:1309.7920 [quant-ph]} \BibitemShut
  {NoStop}%
\bibitem [{\citenamefont {Suchara}\ \emph {et~al.}(2015)\citenamefont
  {Suchara}, \citenamefont {Cross},\ and\ \citenamefont
  {Gambetta}}]{Suchara2015-ll}%
  \BibitemOpen
  \bibfield  {author} {\bibinfo {author} {\bibfnamefont {M.}~\bibnamefont
  {Suchara}}, \bibinfo {author} {\bibfnamefont {A.~W.}\ \bibnamefont {Cross}},
  \ and\ \bibinfo {author} {\bibfnamefont {J.~M.}\ \bibnamefont {Gambetta}},\
  }in\ \href {\doibase 10.1109/ISIT.2015.7282629} {\emph {\bibinfo {booktitle}
  {2015 {IEEE} International Symposium on Information Theory ({ISIT})}}}\
  (\bibinfo {year} {2015})\ pp.\ \bibinfo {pages} {1119--1123}\BibitemShut
  {NoStop}%
\bibitem [{\citenamefont {Weiss}\ \emph {et~al.}(2004)\citenamefont {Weiss},
  \citenamefont {Vala}, \citenamefont {Thapliyal}, \citenamefont {Myrgren},
  \citenamefont {Vazirani},\ and\ \citenamefont {Whaley}}]{Weiss2004-eg}%
  \BibitemOpen
  \bibfield  {author} {\bibinfo {author} {\bibfnamefont {D.~S.}\ \bibnamefont
  {Weiss}}, \bibinfo {author} {\bibfnamefont {J.}~\bibnamefont {Vala}},
  \bibinfo {author} {\bibfnamefont {A.~V.}\ \bibnamefont {Thapliyal}}, \bibinfo
  {author} {\bibfnamefont {S.}~\bibnamefont {Myrgren}}, \bibinfo {author}
  {\bibfnamefont {U.}~\bibnamefont {Vazirani}}, \ and\ \bibinfo {author}
  {\bibfnamefont {K.~B.}\ \bibnamefont {Whaley}},\ }\href {\doibase
  10.1103/PhysRevA.70.040302} {\bibfield  {journal} {\bibinfo  {journal} {Phys.
  Rev. A}\ }\textbf {\bibinfo {volume} {70}},\ \bibinfo {pages} {040302}
  (\bibinfo {year} {2004})}\BibitemShut {NoStop}%
\bibitem [{\citenamefont {Barredo}\ \emph {et~al.}(2016)\citenamefont
  {Barredo}, \citenamefont {de~L{\'e}s{\'e}leuc}, \citenamefont {Lienhard},
  \citenamefont {Lahaye},\ and\ \citenamefont {Browaeys}}]{Barredo2016-bf}%
  \BibitemOpen
  \bibfield  {author} {\bibinfo {author} {\bibfnamefont {D.}~\bibnamefont
  {Barredo}}, \bibinfo {author} {\bibfnamefont {S.}~\bibnamefont
  {de~L{\'e}s{\'e}leuc}}, \bibinfo {author} {\bibfnamefont {V.}~\bibnamefont
  {Lienhard}}, \bibinfo {author} {\bibfnamefont {T.}~\bibnamefont {Lahaye}}, \
  and\ \bibinfo {author} {\bibfnamefont {A.}~\bibnamefont {Browaeys}},\ }\href
  {\doibase 10.1126/science.aah3778} {\bibfield  {journal} {\bibinfo  {journal}
  {Science}\ }\textbf {\bibinfo {volume} {354}},\ \bibinfo {pages} {1021}
  (\bibinfo {year} {2016})},\ \Eprint {http://arxiv.org/abs/1607.03042}
  {arXiv:1607.03042 [quant-ph]} \BibitemShut {NoStop}%
\bibitem [{\citenamefont {Auger}\ \emph {et~al.}(2017)\citenamefont {Auger},
  \citenamefont {Anwar}, \citenamefont {Gimeno-Segovia}, \citenamefont
  {Stace},\ and\ \citenamefont {Browne}}]{Auger2017-di}%
  \BibitemOpen
  \bibfield  {author} {\bibinfo {author} {\bibfnamefont {J.~M.}\ \bibnamefont
  {Auger}}, \bibinfo {author} {\bibfnamefont {H.}~\bibnamefont {Anwar}},
  \bibinfo {author} {\bibfnamefont {M.}~\bibnamefont {Gimeno-Segovia}},
  \bibinfo {author} {\bibfnamefont {T.~M.}\ \bibnamefont {Stace}}, \ and\
  \bibinfo {author} {\bibfnamefont {D.~E.}\ \bibnamefont {Browne}},\ }\href
  {\doibase 10.1103/PhysRevA.96.042316} {\bibfield  {journal} {\bibinfo
  {journal} {Phys. Rev. A}\ }\textbf {\bibinfo {volume} {96}},\ \bibinfo
  {pages} {042316} (\bibinfo {year} {2017})}\BibitemShut {NoStop}%
\bibitem [{\citenamefont {Beterov}\ and\ \citenamefont
  {Saffman}(2015)}]{Beterov2015-zv}%
  \BibitemOpen
  \bibfield  {author} {\bibinfo {author} {\bibfnamefont {I.~I.}\ \bibnamefont
  {Beterov}}\ and\ \bibinfo {author} {\bibfnamefont {M.}~\bibnamefont
  {Saffman}},\ }\href {\doibase 10.1103/PhysRevA.92.042710} {\bibfield
  {journal} {\bibinfo  {journal} {Phys. Rev. A}\ }\textbf {\bibinfo {volume}
  {92}},\ \bibinfo {pages} {042710} (\bibinfo {year} {2015})},\ \Eprint
  {http://arxiv.org/abs/1508.07111} {arXiv:1508.07111 [quant-ph]} \BibitemShut
  {NoStop}%
\bibitem [{\citenamefont {Alberti}\ \emph {et~al.}(2016)\citenamefont
  {Alberti}, \citenamefont {Robens}, \citenamefont {Alt}, \citenamefont
  {Brakhane}, \citenamefont {Karski}, \citenamefont {Reimann}, \citenamefont
  {Widera},\ and\ \citenamefont {Meschede}}]{Alberti2016-su}%
  \BibitemOpen
  \bibfield  {author} {\bibinfo {author} {\bibfnamefont {A.}~\bibnamefont
  {Alberti}}, \bibinfo {author} {\bibfnamefont {C.}~\bibnamefont {Robens}},
  \bibinfo {author} {\bibfnamefont {W.}~\bibnamefont {Alt}}, \bibinfo {author}
  {\bibfnamefont {S.}~\bibnamefont {Brakhane}}, \bibinfo {author}
  {\bibfnamefont {M.}~\bibnamefont {Karski}}, \bibinfo {author} {\bibfnamefont
  {R.}~\bibnamefont {Reimann}}, \bibinfo {author} {\bibfnamefont
  {A.}~\bibnamefont {Widera}}, \ and\ \bibinfo {author} {\bibfnamefont
  {D.}~\bibnamefont {Meschede}},\ }\href {\doibase
  10.1088/1367-2630/18/5/053010} {\bibfield  {journal} {\bibinfo  {journal}
  {New J. Phys.}\ }\textbf {\bibinfo {volume} {18}},\ \bibinfo {pages} {053010}
  (\bibinfo {year} {2016})},\ \Eprint {http://arxiv.org/abs/1512.07329}
  {arXiv:1512.07329 [quant-ph]} \BibitemShut {NoStop}%
\bibitem [{\citenamefont {Martinez~Dorantes}(2016)}]{Martinez_Dorantes2016-ik}%
  \BibitemOpen
  \bibfield  {author} {\bibinfo {author} {\bibfnamefont {M.}~\bibnamefont
  {Martinez~Dorantes}},\ }\emph {\bibinfo {title} {Fast non-destructive
  internal state detection of neutral atoms in optical potentials}},\
  \href@noop {} {Ph.D. thesis},\ \bibinfo  {school} {Universit{\"a}ts-und
  Landesbibliothek Bonn} (\bibinfo {year} {2016})\BibitemShut {NoStop}%
\bibitem [{\citenamefont {Preskill}(1997)}]{Preskill1997-tv}%
  \BibitemOpen
  \bibfield  {author} {\bibinfo {author} {\bibfnamefont {J.}~\bibnamefont
  {Preskill}},\ }\href@noop {} {\  (\bibinfo {year} {1997})},\ \Eprint
  {http://arxiv.org/abs/quant-ph/9712048} {arXiv:quant-ph/9712048 [quant-ph]}
  \BibitemShut {NoStop}%
\bibitem [{\citenamefont {Stephens}(2014)}]{Stephens2014-qc}%
  \BibitemOpen
  \bibfield  {author} {\bibinfo {author} {\bibfnamefont {A.~M.}\ \bibnamefont
  {Stephens}},\ }\href {\doibase 10.1103/PhysRevA.89.022321} {\bibfield
  {journal} {\bibinfo  {journal} {Phys. Rev. A}\ }\textbf {\bibinfo {volume}
  {89}},\ \bibinfo {pages} {022321} (\bibinfo {year} {2014})},\ \Eprint
  {http://arxiv.org/abs/1311.5003} {1311.5003 [quant-ph]} \BibitemShut
  {NoStop}%
\bibitem [{\citenamefont {Wang}\ \emph {et~al.}(2003)\citenamefont {Wang},
  \citenamefont {Harrington},\ and\ \citenamefont {Preskill}}]{Wang2003-mh}%
  \BibitemOpen
  \bibfield  {author} {\bibinfo {author} {\bibfnamefont {C.}~\bibnamefont
  {Wang}}, \bibinfo {author} {\bibfnamefont {J.}~\bibnamefont {Harrington}}, \
  and\ \bibinfo {author} {\bibfnamefont {J.}~\bibnamefont {Preskill}},\ }\href
  {\doibase 10.1016/S0003-4916(02)00019-2} {\bibfield  {journal} {\bibinfo
  {journal} {Ann. Phys.}\ }\textbf {\bibinfo {volume} {303}},\ \bibinfo {pages}
  {31} (\bibinfo {year} {2003})},\ \Eprint
  {http://arxiv.org/abs/quant-ph/0207088} {arXiv:quant-ph/0207088} \BibitemShut
  {NoStop}%
\bibitem [{\citenamefont {Fowler}\ and\ \citenamefont
  {Martinis}(2014)}]{Fowler2014-en}%
  \BibitemOpen
  \bibfield  {author} {\bibinfo {author} {\bibfnamefont {A.~G.}\ \bibnamefont
  {Fowler}}\ and\ \bibinfo {author} {\bibfnamefont {J.~M.}\ \bibnamefont
  {Martinis}},\ }\href {\doibase 10.1103/PhysRevA.89.032316} {\bibfield
  {journal} {\bibinfo  {journal} {Phys. Rev. A}\ }\textbf {\bibinfo {volume}
  {89}},\ \bibinfo {pages} {032316} (\bibinfo {year} {2014})},\ \Eprint
  {http://arxiv.org/abs/1401.2466} {arXiv:1401.2466 [quant-ph]} \BibitemShut
  {NoStop}%
\bibitem [{\citenamefont {Kelly}\ \emph {et~al.}(2015)\citenamefont {Kelly},
  \citenamefont {Barends}, \citenamefont {Fowler}, \citenamefont {Megrant},
  \citenamefont {Jeffrey}, \citenamefont {White}, \citenamefont {Sank},
  \citenamefont {Mutus}, \citenamefont {Campbell}, \citenamefont {Chen},
  \citenamefont {Chen}, \citenamefont {Chiaro}, \citenamefont {Dunsworth},
  \citenamefont {Hoi}, \citenamefont {Neill}, \citenamefont {O'Malley},
  \citenamefont {Quintana}, \citenamefont {Roushan}, \citenamefont
  {Vainsencher}, \citenamefont {Wenner}, \citenamefont {Cleland},\ and\
  \citenamefont {Martinis}}]{Kelly2015-gp}%
  \BibitemOpen
  \bibfield  {author} {\bibinfo {author} {\bibfnamefont {J.}~\bibnamefont
  {Kelly}}, \bibinfo {author} {\bibfnamefont {R.}~\bibnamefont {Barends}},
  \bibinfo {author} {\bibfnamefont {A.~G.}\ \bibnamefont {Fowler}}, \bibinfo
  {author} {\bibfnamefont {A.}~\bibnamefont {Megrant}}, \bibinfo {author}
  {\bibfnamefont {E.}~\bibnamefont {Jeffrey}}, \bibinfo {author} {\bibfnamefont
  {T.~C.}\ \bibnamefont {White}}, \bibinfo {author} {\bibfnamefont
  {D.}~\bibnamefont {Sank}}, \bibinfo {author} {\bibfnamefont {J.~Y.}\
  \bibnamefont {Mutus}}, \bibinfo {author} {\bibfnamefont {B.}~\bibnamefont
  {Campbell}}, \bibinfo {author} {\bibfnamefont {Y.}~\bibnamefont {Chen}},
  \bibinfo {author} {\bibfnamefont {Z.}~\bibnamefont {Chen}}, \bibinfo {author}
  {\bibfnamefont {B.}~\bibnamefont {Chiaro}}, \bibinfo {author} {\bibfnamefont
  {A.}~\bibnamefont {Dunsworth}}, \bibinfo {author} {\bibfnamefont {I.-C.}\
  \bibnamefont {Hoi}}, \bibinfo {author} {\bibfnamefont {C.}~\bibnamefont
  {Neill}}, \bibinfo {author} {\bibfnamefont {P.~J.~J.}\ \bibnamefont
  {O'Malley}}, \bibinfo {author} {\bibfnamefont {C.}~\bibnamefont {Quintana}},
  \bibinfo {author} {\bibfnamefont {P.}~\bibnamefont {Roushan}}, \bibinfo
  {author} {\bibfnamefont {A.}~\bibnamefont {Vainsencher}}, \bibinfo {author}
  {\bibfnamefont {J.}~\bibnamefont {Wenner}}, \bibinfo {author} {\bibfnamefont
  {A.~N.}\ \bibnamefont {Cleland}}, \ and\ \bibinfo {author} {\bibfnamefont
  {J.~M.}\ \bibnamefont {Martinis}},\ }\href {\doibase 10.1038/nature14270}
  {\bibfield  {journal} {\bibinfo  {journal} {Nature}\ }\textbf {\bibinfo
  {volume} {519}},\ \bibinfo {pages} {66} (\bibinfo {year} {2015})}\BibitemShut
  {NoStop}%
\bibitem [{\citenamefont {Barends}\ \emph {et~al.}(2014)\citenamefont
  {Barends}, \citenamefont {Kelly}, \citenamefont {Megrant}, \citenamefont
  {Veitia}, \citenamefont {Sank}, \citenamefont {Jeffrey}, \citenamefont
  {White}, \citenamefont {Mutus}, \citenamefont {Fowler}, \citenamefont
  {Campbell}, \citenamefont {Chen}, \citenamefont {Chen}, \citenamefont
  {Chiaro}, \citenamefont {Dunsworth}, \citenamefont {Neill}, \citenamefont
  {O'Malley}, \citenamefont {Roushan}, \citenamefont {Vainsencher},
  \citenamefont {Wenner}, \citenamefont {Korotkov}, \citenamefont {Cleland},\
  and\ \citenamefont {Martinis}}]{Barends2014-tf}%
  \BibitemOpen
  \bibfield  {author} {\bibinfo {author} {\bibfnamefont {R.}~\bibnamefont
  {Barends}}, \bibinfo {author} {\bibfnamefont {J.}~\bibnamefont {Kelly}},
  \bibinfo {author} {\bibfnamefont {A.}~\bibnamefont {Megrant}}, \bibinfo
  {author} {\bibfnamefont {A.}~\bibnamefont {Veitia}}, \bibinfo {author}
  {\bibfnamefont {D.}~\bibnamefont {Sank}}, \bibinfo {author} {\bibfnamefont
  {E.}~\bibnamefont {Jeffrey}}, \bibinfo {author} {\bibfnamefont {T.~C.}\
  \bibnamefont {White}}, \bibinfo {author} {\bibfnamefont {J.}~\bibnamefont
  {Mutus}}, \bibinfo {author} {\bibfnamefont {A.~G.}\ \bibnamefont {Fowler}},
  \bibinfo {author} {\bibfnamefont {B.}~\bibnamefont {Campbell}}, \bibinfo
  {author} {\bibfnamefont {Y.}~\bibnamefont {Chen}}, \bibinfo {author}
  {\bibfnamefont {Z.}~\bibnamefont {Chen}}, \bibinfo {author} {\bibfnamefont
  {B.}~\bibnamefont {Chiaro}}, \bibinfo {author} {\bibfnamefont
  {A.}~\bibnamefont {Dunsworth}}, \bibinfo {author} {\bibfnamefont
  {C.}~\bibnamefont {Neill}}, \bibinfo {author} {\bibfnamefont
  {P.}~\bibnamefont {O'Malley}}, \bibinfo {author} {\bibfnamefont
  {P.}~\bibnamefont {Roushan}}, \bibinfo {author} {\bibfnamefont
  {A.}~\bibnamefont {Vainsencher}}, \bibinfo {author} {\bibfnamefont
  {J.}~\bibnamefont {Wenner}}, \bibinfo {author} {\bibfnamefont {A.~N.}\
  \bibnamefont {Korotkov}}, \bibinfo {author} {\bibfnamefont {A.~N.}\
  \bibnamefont {Cleland}}, \ and\ \bibinfo {author} {\bibfnamefont {J.~M.}\
  \bibnamefont {Martinis}},\ }\href {\doibase 10.1038/nature13171} {\bibfield
  {journal} {\bibinfo  {journal} {Nature}\ }\textbf {\bibinfo {volume} {508}},\
  \bibinfo {pages} {500} (\bibinfo {year} {2014})},\ \Eprint
  {http://arxiv.org/abs/1402.4848} {arXiv:1402.4848 [quant-ph]} \BibitemShut
  {NoStop}%
\bibitem [{\citenamefont {Ballance}\ \emph {et~al.}(2016)\citenamefont
  {Ballance}, \citenamefont {Harty}, \citenamefont {Linke}, \citenamefont
  {Sepiol},\ and\ \citenamefont {Lucas}}]{Ballance2016-hu}%
  \BibitemOpen
  \bibfield  {author} {\bibinfo {author} {\bibfnamefont {C.~J.}\ \bibnamefont
  {Ballance}}, \bibinfo {author} {\bibfnamefont {T.~P.}\ \bibnamefont {Harty}},
  \bibinfo {author} {\bibfnamefont {N.~M.}\ \bibnamefont {Linke}}, \bibinfo
  {author} {\bibfnamefont {M.~A.}\ \bibnamefont {Sepiol}}, \ and\ \bibinfo
  {author} {\bibfnamefont {D.~M.}\ \bibnamefont {Lucas}},\ }\href {\doibase
  10.1103/PhysRevLett.117.060504} {\bibfield  {journal} {\bibinfo  {journal}
  {Phys. Rev. Lett.}\ }\textbf {\bibinfo {volume} {117}},\ \bibinfo {pages}
  {060504} (\bibinfo {year} {2016})},\ \Eprint
  {http://arxiv.org/abs/1512.04600} {arXiv:1512.04600 [quant-ph]} \BibitemShut
  {NoStop}%
\bibitem [{\citenamefont {Kolmogorov}(2009)}]{Kolmogorov2009-md}%
  \BibitemOpen
  \bibfield  {author} {\bibinfo {author} {\bibfnamefont {V.}~\bibnamefont
  {Kolmogorov}},\ }\href {\doibase 10.1007/s12532-009-0002-8} {\bibfield
  {journal} {\bibinfo  {journal} {Math. Program. Comput.}\ }\textbf {\bibinfo
  {volume} {1}},\ \bibinfo {pages} {43} (\bibinfo {year} {2009})}\BibitemShut
  {NoStop}%
\end{thebibliography}
\end{document}